\documentclass[review]{elsarticle}

\usepackage{graphicx,epsfig}
\usepackage{amsmath}
\usepackage{multirow}
\usepackage{color}
\usepackage[left]{lineno}

\def\beq{\begin{equation}}
\def\eeq{\end{equation}}
\def\bea{\begin{eqnarray}}
\def\eea{\end{eqnarray}}
\def\nn{\nonumber}

\def\d{\mbox{d}}

\def\etal{\mbox{\it et al.} }

\def\blu{\color{blue}}

\def\blu{}

\def\xa{r_b}
\def\xb{{l}_{bc}}
\def\xc{r_a}
\def\xd{{l}_{ac}}
\def\xe{r_d}
\def\xf{Re^{1/3}}
\def\xg{Re^{2/3}}

\begin{document}

\noindent {\it This manuscript has been accepted for publication in Journal of the Acoustical Society of America}

\begin{center}
\textbf{\Large
A one-dimensional flow model enhanced by machine learning for simulation of vocal fold vibration}
\end{center}


\begin{center}
Zheng Li$^a$,
Ye Chen$^a$,
Siyuan Chang$^a$\footnote{Current affiliation:Corning Inc., Corning, NY},
Bernard Rousseau$^b$,
Haoxiang Luo$^a$\footnote{Corresponding author: \texttt{haoxiang.luo@vanderbilt.edu} (E-mail), +1-615-322-2079 (Tel)},
\\
\textit{$^a$Department of Mechanical Engineering, Vanderbilt University,
   2301 Vanderbilt Place, Nashville, TN 37235-1592}
\\
\textit{$^b$Department of Communication Science and Disorders,
University of Pittsburgh}

\end{center}

\centerline{\bf Abstract}


We describe a one-dimensional (1D) unsteady and viscous flow model that is derived from the momentum and mass conservation equations, and to enhance this physics-based model, we use a machine learning approach to determine the unknown modeling parameters.  Specifically, we first construct an idealized larynx model and perform ten cases of three-dimensional (3D) fluid--structure interaction (FSI) simulations.  The flow data are then extracted to train the 1D flow model using a sparse identification approach for nonlinear dynamical systems.  As a result of training, we obtain the analytical expressions for the entrance effect and pressure loss in the glottis, which are then incorporated in the flow model to conveniently handle different glottal shapes due to vocal fold vibration.  We apply the enhanced 1D flow model in the FSI simulation of both idealized vocal fold geometries and subject-specific anatomical geometries reconstructed from the MRI images of rabbits' larynges.  The 1D flow model is evaluated in both of these setups and is shown to have robust performance.  Therefore, it provides a fast simulation tool superior to the previous 1D models.



\section{Introduction}

Computational modeling of fluid--structure interaction (FSI) for vocal fold vibration is useful as it may provide a computer based tool for clinical management of voice disorders,  e.g., surgical planning for vocal fold paralysis~\cite{mittal2011toward}.  Despite that the underlying physical principle of vocal fold vibration is straightforward and can be modeled simply using lumped-mass models, high-fidelity modeling to simulate details of the tissue's dynamic deformation is still very challenging, especially if patient-specific features should be simulated for the purpose of developing modeling tools that can capture differences in the laryngeal anatomy and tissue properties of individuals.

With tremendous growth of the computer power and improvement of the modeling approach, computational models for the FSI of vocal fold vibration have been advanced substantially in recent years.  These physics-based models typically couple a 2D or 3D glottal airflow model and a finite-element representation of the vocal fold tissue~\cite{alipour2000finite,hunter2004three,thomson2005aerodynamic,cook2007sensitivity,luo2008immersed,luo2009analysis,shurtz2013influence,chang2013role,zhang2017effect, yang2017fully, valavsek2019application,sadeghi2019computational}.  In addition, they adopt increasingly high resolution and have provided insightful fundamental understanding about this FSI system such as eigenmodes~\cite{alipour2000finite}, mechanics of posturing~\cite{hunter2004three}, sensitivity to the geometry and material properties~\cite{cook2007sensitivity,shurtz2013influence,chang2013role,zhang2017effect}, vortex flow and pressure on the vocal fold surface~\cite{luo2008immersed,luo2009analysis,sadeghi2019computational}, energy transfer~\cite{thomson2005aerodynamic,luo2009analysis}, and acoustic wave propagation~\cite{yang2017fully, valavsek2019application}. In these references, the vocal fold was often represented by a schematic that captures only the overall characteristics of the laryngeal geometry. Such models are obviously not sufficient for subject-specific representation.  
 Medical imaging technologies, e.g., computed tomography (CT) and magnetic resonance imaging (MRI), allows the computational models to incorporate more realistic or even subject-specific laryngeal geometries.  These imaging tools may provide detailed 3D anatomy of the larynx, as well as the interior structure of the tissue~\cite{madruga2003distribution,pickup2010flow, wu2016parametric, novaleski2016nonstimulated}.  Coupled with a 3D airflow solver in FSI simulation, the anatomical models of the larynx represent a significant step toward patient-specific modeling of vocal fold vibration, which {\blu may be} needed for clinical care of voices of individual patients. In recent years, such patient-specific models have been developed to study vocal fold vibration by {\blu several researchers~\cite{mittal2013fluid,xue2014subject,chang2015subject}}.

On the other hand, the unknown tissue properties for individual subjects, which can not be identified from current imaging technologies, limit the application of patient-specific modeling.  There have been prior efforts to derive the vocal fold tissue properties using finite-element method (FEM) based models combined with experimental tests~\cite{schmidt2011material,schmidt2013material,alipour2013phonatory}, but they were limited to {\it ex-vivo} conditions.  Although it is probable to determine the {\it in-vivo}, subject-specific tissue properties by running the high-fidelity FSI simulations and solving an inverse problem,  assuming that all other aspects in the FSI model match the corresponding {\it in-vivo} experiment (e.g., the anatomy and boundary conditions), such an approach is still not practical since the 3D airflow simulation is too expensive even with high-performance parallel computing.  More practically, one could use a simplified flow model, coupled with a realistic FEM representation of the vocal fold for the FSI simulation, to determine the elastic properties with much lower computational cost.  Such efforts have been made recently in a few studies~\cite{chang2015subject,dollinger2017biomechanical,hadwin2019bayesian,zhang2020estimation}.  In practice, this kind of simplified FSI models could also be combined with the high-fidelity models to increase the overall modeling accuracy~\cite{chang2015subject}.

Bernoulli based flow equations have long been used for vocal fold vibration. Decker and Thomson~\cite{decker2007computational} compared the Bernoulli equation with the Navier--Stokes equation for simulation of vocal fold vibration, and they found that the Bernoulli models could be highly dependent on the {\it ad hoc} assumption of the flow separation in the glottis.   Chang~\cite{chang2016computational} also reached a similar conclusion, and they found the Bernoulli model may lead to a significantly different vibration mode of the vocal fold from a full Navier--Stokes model.  In general, the main limitations in the Bernoulli principle are due to the assumption of the ideal flow in the glottis and a priori unknown location of flow separation. To address the limitations of the Bernoulli equation, in our recent works~\cite{li2019reduced,li2019reduced2} we developed a 1D momentum equation based flow model that was originally designed to solve separated flow in the collapsible tube~\cite{cancelli1985separated,anderson2013implementation} and was only recently introduced for  modeling of vocal fold dynamics~\cite{vasudevan2017fast}. In this model, we have included the effect of pressure loss that is caused by the flow separation and the viscous effect. Furthermore, we have included an entrance effect, which is due to an inertial flow entering the glottis from a rapidly converging shape in the subglottal region.  This 1D flow model was coupled with the 3D tissue model for FSI simulation of both idealized and anatomical vocal fold geometries from rabbits.  In the idealized vocal fold cases, the reduced FSI model achieved consistent results with those from the full 3D FSI model for different medial vocal fold thicknesses, subglottal pressures, tissue models, and tissue stiffness properties.  In the anatomical models, vibration results from the reduced FSI model agreed well with the  experimental data of the evoked {\it in-vivo} rabbit phonation~\cite{li2019reduced,li2019reduced2}. 

However, the pressure loss and entrance effect in the glottal airflow depend on the overall shape of the glottis as well as its instantaneous deformation during vibration.  In addition, the Reynolds number plays a significant role.  Since no analytical expression exist to describe these effects precisely, in the previous work~\cite{li2019reduced,li2019reduced2} we chose to use constant parameters  based on the knowledge learned from 3D flow simulations. These parameters may need adjustment depending on specific vocal fold geometry, e.g., a large or small medial thickness, in order to achieve good accuracy. This limitation reduces generalization of the flow model.  To overcome the limitation, we set variable parameters for the pressure loss and entrance effect and seek to express them in a functional form that is convenient to use.  

Machine learning techniques, which have gained popularity in fluid mechanics in recent years~\cite{brunton2019machine}, provide a useful approach to help determine the  functional form of a physical effect, especially when general characteristics of such effect have been understood.   This feature applies well to the situation we are considering.  Here we use the 1D Navier--Stokes equation to describe the flow physics and leave only the undescribed effects to be determined empirically.  These effects, i.e., the pressure loss and entrance effect in a converging-diverging glottis, can be expressed as functions of only a few dimensionless parameters such as the instantaneous shape of the glottis and the Reynolds number of the flow. Once these functions are determined using a training data set through machine learning, they can be incorporated into the 1D model for flow simulation of arbitrary glottal shapes. Recently, there have been other studies that also incorporated machine learning into  modeling of vocal fold dynamics~\cite{zhang2020estimation,gomez2018laryngeal,zhang2020deep}. Compared to those studies, the present work takes advantage of the physics-based model, i.e., the unsteady viscous Navier-Stokes equation, albeit within the limitations of 1D, as much as possible and only resort to machine learning for the remaining undescribed effects that includes the entrance effect.

In fluid mechanics, many machine learning techniques have been developed to identify the governing equation directly from data ~\cite{brunton2016discovering,raissi2017machine,berg2019data,brunton2019machine}. Here, we adopt the sparse identification of nonlinear dynamical (SINDy) systems~\cite{brunton2016discovering}, which is a regression algorithm suitable for the physical systems having only a few relevant terms to define the dynamics. Specifically, we will use the SINDy method to identify the pressure loss and entrance effect at locations near the glottal exit and express them as polynomial functions of the Reynolds number, the channel length, and the convergent or divergent ratios of the glottis. To generate the training data for machine learning, we will use 3D FSI simulations of an idealized vocal fold with different medial thicknesses, stiffness properties, and subglottal pressures.  Additional 3D simulation cases will be performed for validation of the machine learning. To further assess the performance of the new flow model, we will compare other quantities such as the pressure distribution, flow rate, and vibratory characteristics in the idealized model against the 3D simulation.  Furthermore, we will apply the new model to FSI simulation of anatomical vocal fold geometries that are based on excised rabbit larynges.  The simulation results will be compared with high-speed video data from {\it in-vivo} phonation of the same larynx samples prior to the excision.  The descriptions of the model setup, machine learning procedure, and results from machine learning and FSI simulations are provided in the following sections.

\section{The flow model and training case setup}

\subsection{The one-dimensional viscous flow model}
\label{sec-1dflow}

\begin{figure}
    \centering
        \epsfig{file=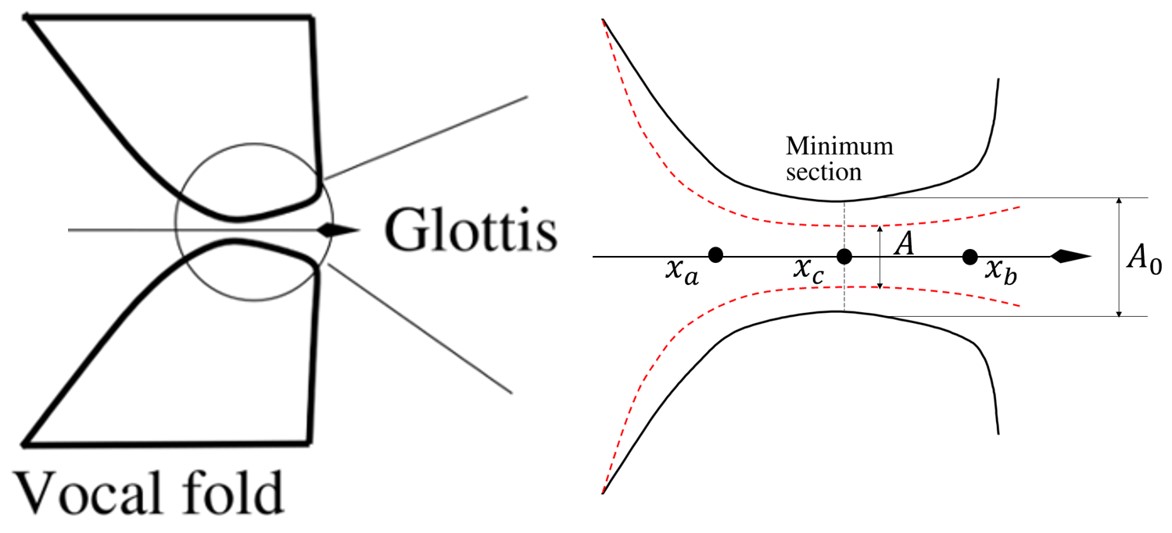,width=4.0in}
    
    \caption{Schematic of airflow entering the glottis that generally has a converging-diverging shape.  The flow experiences loss of energy and also a vena contracta effect, in which the effective area, $A$, may be smaller than the actual cross section area, $A_0$. The red-dashed lines indicate the boundary layer.}
    
    \label{fig:ent}
\end{figure}

A 2D schematic of the geometry in the transverse plane of the larynx is shown in Fig.~\ref{fig:ent}, where the  glottis is depicted as a converging-diverging channel.  To facilitate our discussion, we use $x_a$ and $x_b$ to mark the locations of the glottal inlet and exit, respectively, and $x_c$ the location of the minimal cross-section area in the glottis.  Note that $x_c$ varies between $x_a$ and $x_b$ when the vocal fold is vibrating, and it could coincide with $x_b$ such that the glottis is purely convergent.  In practice, the location of $x_b$ is straightforward to choose as the cross section typically experiences a sudden expansion at the glottal exit.  On the other hand, the location of $x_a$
sometimes is not obvious since the subglottal region may narrow down gradually rather than abruptly. We point out that from our tests, the present flow model is not sensitive to the inlet location if the glottis does not have a clear entrance location, in which case an approximate choice of $x_a$ would be sufficient.  This is because the entrance effect of a gradually convergent section is small anyway, and in addition, the pressure loss primarily takes effect in the diverging section in the present model and has little dependence on the inlet location.

When the air flows through the glottis during vibration, the pressure generally decreases before the minimal area section at $x_c$ due to the Bernoulli effect. After this section, the pressure would increase along with the area expansion.  However, the pressure will not recover to its full extent because of the possible separation in  the divergent section and also the viscous effect in the entire glottis.  Therefore, accounting for the pressure loss in the flow model will help overcome  limitations of the Bernoulli equation.  Considering the mass and momentum conservation  equations, Cancelli and Pedley~\cite{cancelli1985separated} developed a 1D flow model to describe a collapsible tube.  In the momentum equation, they included the viscous loss and separation effects.  To generalize the pressure loss, we combine the viscous loss and separation effects as one single loss term represented by the shear stress, $\tau$, in the following equation
\begin{eqnarray}
\frac{\partial A}{\partial t} +\frac{\partial Au}{\partial x} &=& 0
\nonumber\\
\rho\frac{\partial u}{\partial t}+\rho u\frac{\partial u}{\partial x} &=& -\frac{\partial p}{\partial x} + \frac{\partial \tau}{\partial x}
\label{e:mom}
\end{eqnarray}
where $\rho$, $u$ and $p$ are respectively the density, velocity, and pressure, and $A$ is the effective area of the cross section.  We will discuss the calculation of the shear stress $\tau(x)$ later using a machine learning approach.  For the boundary conditions of the 1D flow, we set a specified subglottal pressure, $P_{sub}$, and the pressure at the glottal exit, $P_{e}$.  Eq.~\eqref{e:mom} represents a nonlinear boundary value problem and can be solved using a shooting method once we have an expression for $\tau$.  Its numerical procedure was  described in Li et al.~\cite{li2019reduced}.

Beside the loss term, the vena contracta effect of flow entering an expansion was introduced in our flow model~\cite{li2019reduced,li2019reduced2}.  In particular, as the air flows into the glottis, and especially the diverging section, it tends to focus to the center under its inertia, rather than following the exact shape of the channel.  Thus, we use the effective cross section area, $A$, in Eq.~\eqref{e:mom} for the mass conservation equation. This area is smaller than the actual cross sectional area, $A_0$, as illustrated in Fig.~\ref{fig:ent}.  Without such an entrance effect, the {\blu negative pressure (gage pressure) } at the minimum section could be overestimated, leading to inaccurate pressure load on the vocal fold surface.  To calculate the effective area $A$, we introduce a correctional coefficient, $\alpha(x)$, so that 
\begin{equation}
A(x)=\alpha(x) A_0(x).
\label{e:AA0}
\end{equation}
Note that $\alpha$ is a function of the streamwise location, $x$.

In our previous work~\cite{li2019reduced,li2019reduced2}, we estimated $\alpha(x)$ based on the 3D simulation of the FSI problem by calculating it from $\alpha(x) = u_{avg}/u$, where $u_{avg}$ is the average streamwise velocity in the cross section and $u$ is the maximum streamwise velocity.  We further assumed a quadratic function form for $\alpha(x)$ with a single free parameter to be determined through machine learning, as will be discussed in next section. The quadratic function represents narrowing down of the effective area due to the growth of the boundary layer along the glottis.

\subsection{Input variables for machine learning} 
\label{s:ml-in}

\begin{figure}
\centering
    
\hskip-3.0in (a) \hskip3.0in (b)

\includegraphics[width=2.0in]{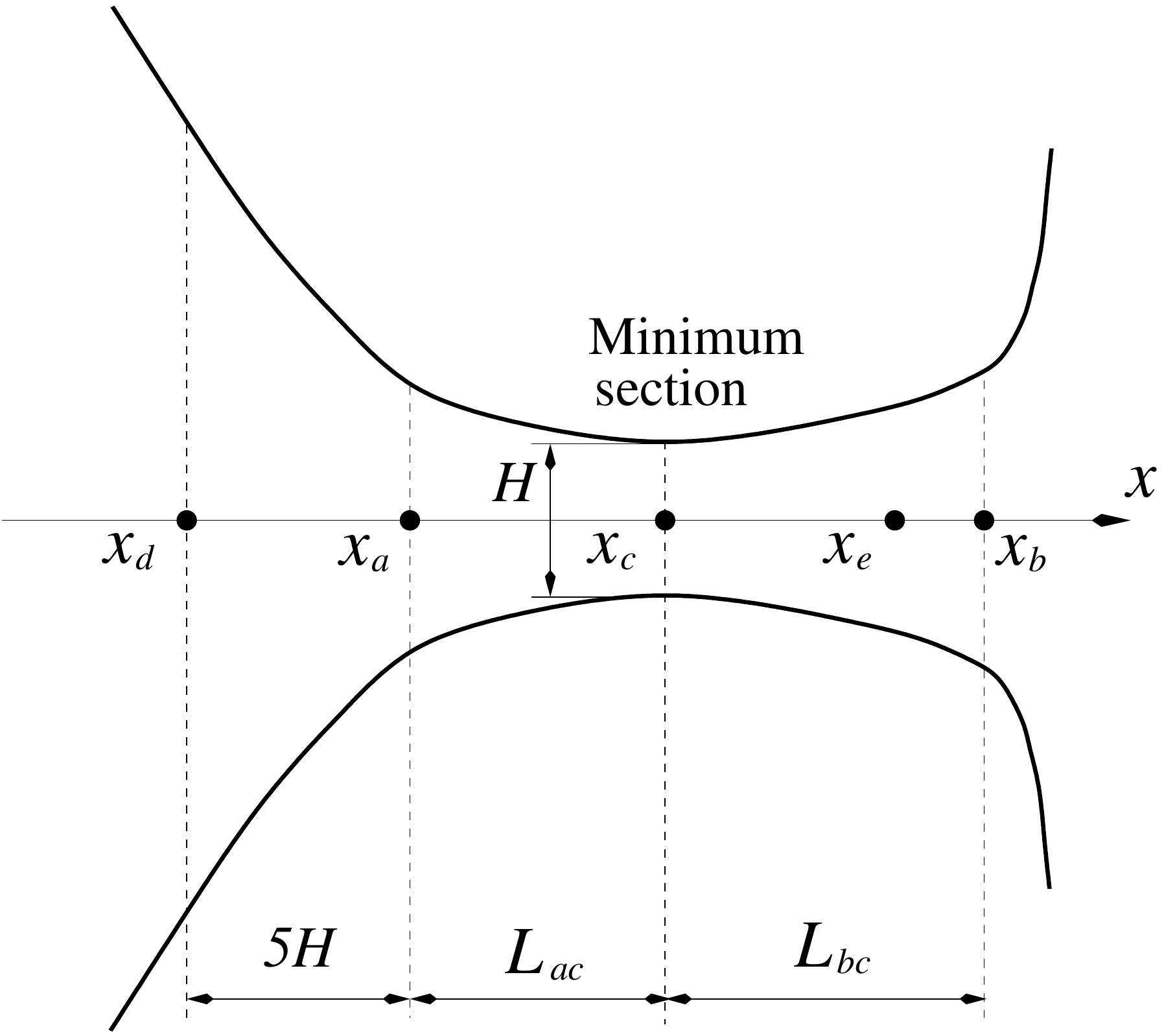}
\hskip0.5in
\includegraphics[width=2.0in]{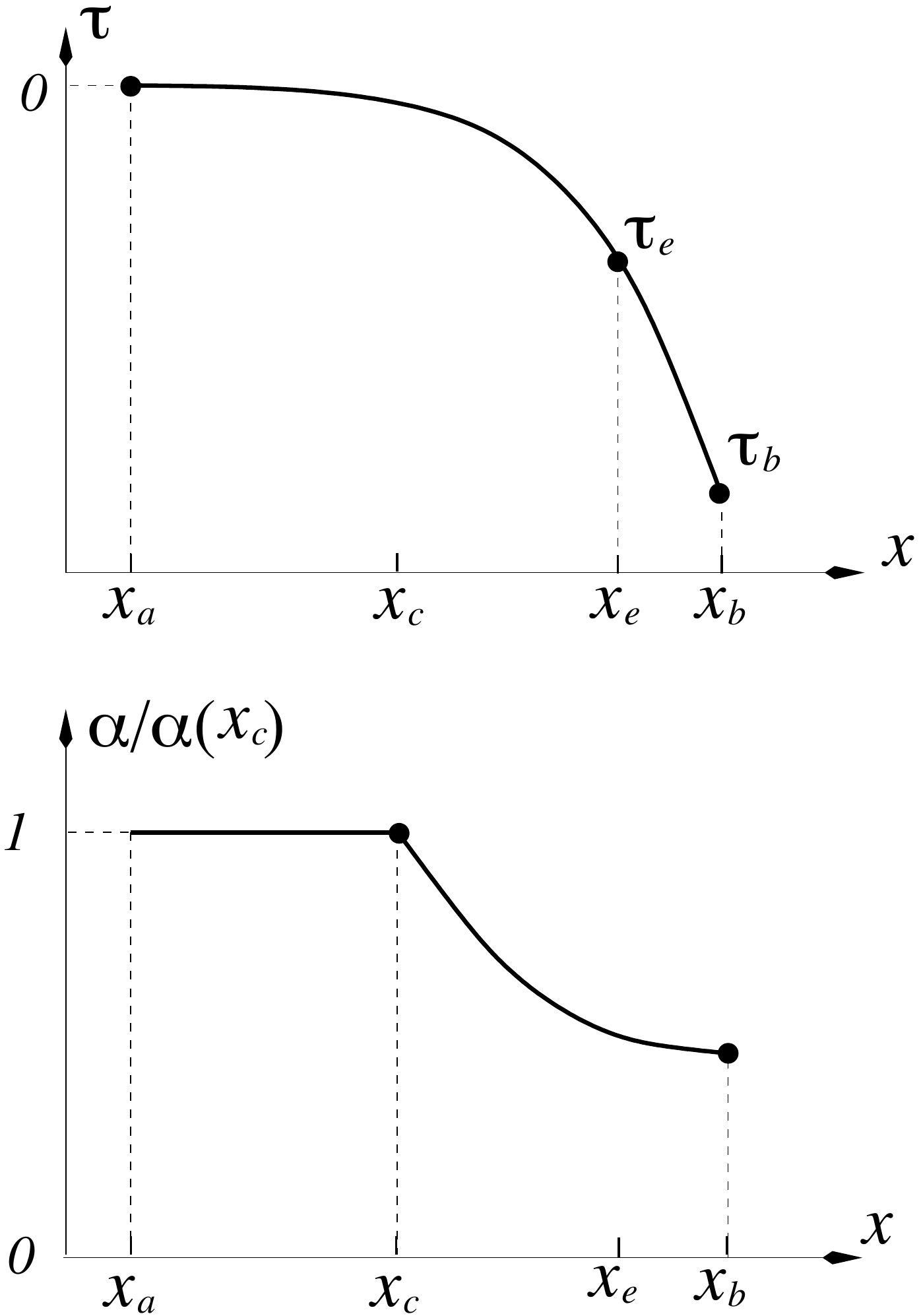}
    
\caption{(a) Geometric description of the glottis used in the machine learning. (b) Functions of the pressure loss, $\tau(x)$, and the area correction coefficient, $\alpha(x)$, along the glottis.}
\label{f:ml-geo}
\end{figure}

To outline the flow model for machine learning, we use a simplified but characteristic geometry of the glottis and define the input--output variables for the machine learning module.  Fig.~\ref{f:ml-geo}{(a)} shows a 2D schematic of the glottis.  We define five locations along the flow, which are: 1) the glottal inlet $x_a$, 2) the glottal exit $x_b$, 3) the narrowest area location $x_c$, 4) a point in the subglottal region $x_d$, and 5) an intermediate location in the divergent section, $x_e$.  The average gap width at the narrowest section is denoted as $H$, so $H=A(x_c)/L$, where $A(x_c)$ is the cross section area at $x_c$ and $L$ is the longitudinal length of the glottis.  
These locations, $x_a$, $x_b$, $x_c$, and the corresponding cross-sectional area, $A(x_a)$, $A(x_b)$, and $A(x_c)$,  describe the overall converging-diverging shape of the glottis.  In addition to these variables, $x_d$ and $A(x_d)$ are used to describe the slope of the subglottal region, which is useful in measuring the extent that the flow is focused when entering the glottis. Previous study has shown that the geometry of the glottal entrance has a significant influence on the intraglottal pressure distribution~\cite{li2012effect}. Here we set $x_d$ at a distance of $5H$ from $x_a$ to capture the slope of the subglottal shape.  Furthermore, the intermediate point $x_e$ is added so that the pressure loss at this location will be determined as one output variable as described in next section.   We set $x_e$ to be closer to $x_b$, with a distance ratio of 3:1 between $|x_c-x_e|$ and $|x_b-x_e|$, since the pressure loss increases more quickly near the glottal exit, as illustrated in Fig.~\ref{f:ml-geo}(b).
It is worth pointing out that the exact location of $x_d$ and $x_e$ are not crucial, as $x_d$ is used to calculate the subglottal slope of the vocal fold
and $x_e$ is used to provide another data point for the pressure loss estimate.

In terms of nondimensional parameters, the geometric variables along with the Reynolds number are defined as below,
\begin{eqnarray}
      \xa&=&A(x_b)/A(x_c) 
      \nn\\
      \xb&=&L_{bc}/H
      \nn\\
      \xc&=&A(x_a)/A(x_c)
      \nn \\
      \xd&=&L_{ac}/H
      \nn\\
      \xe&=&A(x_d)/A(x_a)
      \nn\\
      \xf&=& \left[ \rho\, u_c\, H \over \mu \right]^{1/3} 
\label{e:ml-in}
 \end{eqnarray}
Among these six variables, $\xa$, $\xc$, and $\xe$ are the area ratios, $\xb$ and $\xd$ are the normalized distances. The Reynolds number is defined using the velocity at the narrowest section, $u_c$, and is reduced to the power $1/3$ so that this variable is at a similar order of magnitude as the other five.

\subsection{Output variables for machine learning}
\label{s:ml-out}

To determine the pressure loss, or the shear stress term in Eq.~\eqref{e:mom}, we assume that $\tau=0$ and $\d\tau/\d x=0$ at the glottal inlet $x_a$, $\tau=\tau_e$ at the intermediate point $x_e$, and $\tau=\tau_b$ at the exit $x_b$.  The two unknown variables, $\tau_e$ and $\tau_b$, will be determined using machine learning as functions of the six input variables described in Eq.~\eqref{e:ml-in}.   Once $\tau_e$ and $\tau_b$ are determined from machine learning, we assume a cubic distribution for $\tau(x)$ from the glottal inlet $x_a$ to the exit $x_b$ as shown in Fig.~\ref{f:ml-geo}(b).  This assumption is made based on observation of general characteristics of the pressure loss from our 3D flow simulations~\cite{li2019reduced2}.  Note that a higher order distribution is also possible using the same strategy, if more output variables are used from machine learning.  


For the entrance effect, we need to determine the area correction coefficient, $\alpha(x)$, along the glottis.  Similar to our previous publications~\cite{li2019reduced,li2019reduced2}, we assume a quadratic distribution for $\alpha(x)$ between $x_c$ and $x_b$.  However, in the present study we will only need to determine $\alpha/\alpha(x_c)$ since only the relative area ratio is needed when solving Eq.~\eqref{e:mom}.  Furthermore,  $\alpha(x)$ is assumed to have zero derivative at $x_b$.  Therefore, we will only need to determine  $\alpha(x_b)/\alpha(x_c)$ through machine learning.
In summary, there are three output variables for the machine learning process, which are $\tau_e$, $\tau_b$, and $\alpha(x_b)/\alpha(x_c)$.

\subsection{The SINDy method for machine learning}
\label{s:ml-sindy}

For machine learning, we use sparse identification of nonlinear dynamical systems (SINDy)~\cite{brunton2016discovering}, which is a data regression approach to discover governing equations for nonlinear dynamical systems including fluid flows. In particular, SINDy uses
sparse regression to determine the fewest terms in the dynamic
governing equations required to accurately represent the data, and this
results in parsimonious models that balance accuracy with model
complexity to avoid overfitting~\cite{brunton2016discovering}.  The key assumption in this approach is that for many systems of interest, the governing equation consists of only a few terms, making it sparse in the space of possible functions~\cite{brunton2016discovering}. Using training data that will be described in next section, SINDy can determine a generic output variable, $f$, as a polynomial function of the six input variables defined in Eq.~\eqref{e:ml-in}, i.e.,
\begin{equation}
f = f(\xa,\,\xb,\,\xc,\,\xd,\,\xe,\,\xf,\,\xa^2,\,\xa\xb,\,\cdots,\, \xg, \cdots)
\label{e:ml-pol}
\end{equation}
The software package of SINDy in Matlab is freely available by the authors~\cite{brunton2016discovering} and is used here for our study.  For our study, only the terms up to the third order are retained in this polynomial function.

\subsection{Setup of the 3D FSI model and data generation}
\label{s:3dmod}

\begin{figure}
    \centering
        \epsfig{file=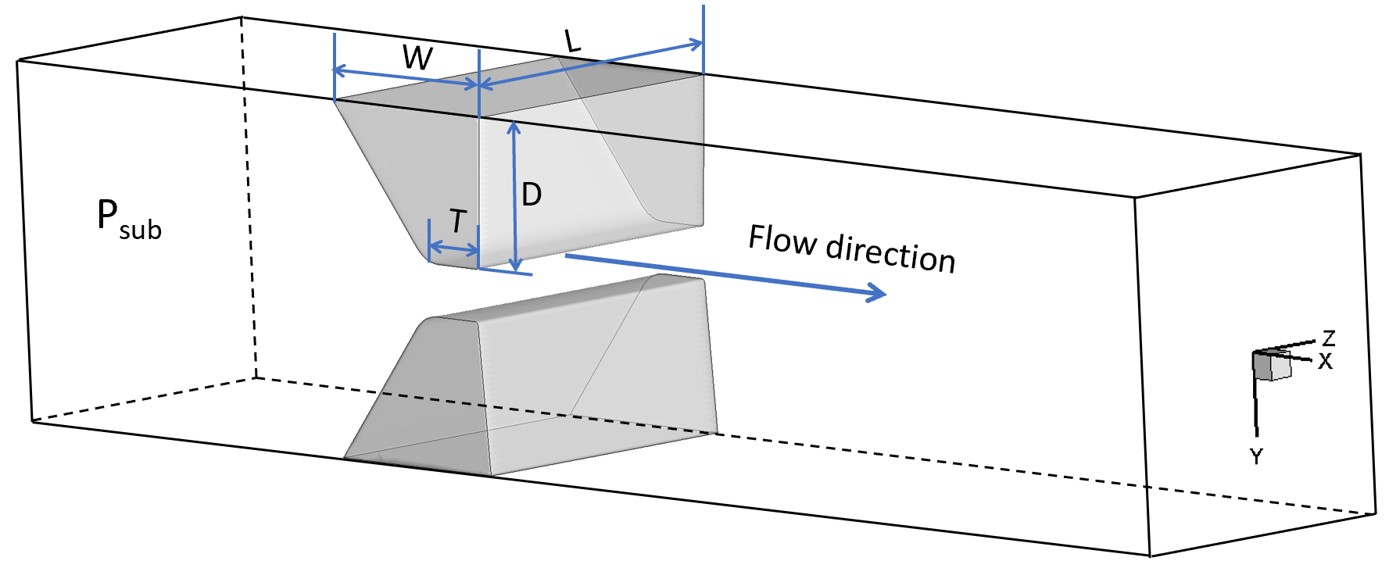,width=5in}
    \caption{The vocal fold model and computational domain used for 3D FSI simulation and data generation.}
    \label{f:3Dmodel}
\end{figure}

\begin{table}
\begin{center}
    \caption{FSI cases setup for data generation. Cases 1-6 are used for training, and Cases 7-10 for additional validation.}
\begin{tabular}{ |c|c|c|c|c|c|c|c|c|c|c| } 
\hline
Case & 1 & 2 & 3 & 4 & 5 & 6 & 7 & 8 & 9 & 10\\
\hline
$T$ (mm)& \multicolumn{3}{| c |}{1.75}& \multicolumn{3}{| c |}{3.50} &\multicolumn{2}{| c |}{1.75} & \multicolumn{2}{| c |}{3.50} \\
\hline
$P_{sub}$ (kPa) & \multicolumn{6}{| c |}{1.00}& 1.25 & 0.75 
& 1.25 & 0.75\\
\hline
$\alpha_{10}$ (kPa) &2.29& 2.58 & 9.16 & 2.29 & 4.58 & 9.16 &\multicolumn{4}{| c |}{2.29}\\
\hline
$\alpha_{01}$ (kPa) & 0.25 & 0.50 & 1.00 & 0.25 & 0.50 & 1.00 & \multicolumn{4}{| c |}{0.25}\\
\hline
\end{tabular}
\label{t:ml-cases}
\end{center}
\end{table}

To generate training data for machine learning, we use a previous 3D setup of the FSI of an idealized vocal fold geometry as illustrated in Fig.~\ref{f:3Dmodel}. This model has been described in our previous publications~\cite{chang2013role,li2019reduced,li2019reduced2} and is only briefly summarized here.  The airflow is driven by a constant subglottal pressure $P_{sub}$, and the outlet has a reference pressure of $P_{out}=0$ kPa for all the cases in consideration. The air is assumed to be incompressible and is governed by the viscous Navier-Stokes equation.  A pair of vocal fold bands are placed symmetrically in the channel, whose length, width, and depth are $L=20$ mm, $W=13$ mm, and $D=10$ mm, respectively.  The medial thickness, $T$, has significant effects on the flow and the vocal fold vibration~\cite{chang2016computational}, as the medial surfaces are the primary loading surfaces for the sustained vibration.  The vocal fold here is assumed to be isotropic, homogeneous, and is governed by a hyperelastic, two-parameter Mooney-Rivlin model. The strain energy density function for this model is given as
\begin{equation}
{\mathcal W}=\alpha_{10}(\overline{I}_1-3)+\alpha_{01}(\overline{I}_2-3)+K/2(J-1)^2
\label{e:mr energy}
\end{equation}
where $K$ represents the bulk modulus, $\alpha_{10}$ and $\alpha_{01}$  are material constants related to the distortional response, and $J = \det(F)$ with $F$ standing for the deformation gradient.  In addition, $\overline{I}_1$  and  $\overline{I}_2$ are invariants based on $J$ and the principal stretches of the deformation gradient. Further detail of this model for the vocal fold can be found in our group's previous work~\cite{tian2014fluid}.  Anisotropic tissue behavior or a multi-layer structure as proposed in many previous works~\cite{alipour2000finite,luo2009analysis,zhang2017effect} would be better representation of the real tissue of vocal fold.  However, we only need characteristic vocal fold deformation and corresponding flow data for the training purpose; thus, the specific material model for the vocal fold tissue is not essential in this study. 

To solve the 3D FSI, an in-house immersed-boundary method is employed for the flow simulation, while the tissue deformation is solved with a finite-element method~\cite{tian2014fluid}. In total, we solved ten simulation cases after a careful mesh  independence study~\cite{li2019reduced,li2019reduced2}. Table~\ref{t:ml-cases} shows the details for all the test cases, which contains variations in these parameters: the medial thickness $T$, the subglottal pressure $P_{sub}$, and the material stiffness constants $\alpha_{10}$ and $\alpha_{01}$.  The tissue density is $\rho_s = 1040$ kg/m$^3$ and mass damping is 0.05 s$^{-1}$ in all the cases. The air density is $\rho=1.13$ kg/m$^3$. Thus, the characteristic intraglottal velocity is $V = \sqrt{2 (P_0-P_{out})/\rho}=42.1$ m/s. We define the jet Reynolds number using $Re_J=\rho V d/\mu$, where $d\sim 1$ mm is the characteristic glottal gap during opening phase and $\mu$ is the air viscosity.  In the current study, we set $Re_J=210$.

Cases 1-6 are utilized in SINDy as training data, which have the same $P_{sub}$ but two different medial thicknesses and three stiffness constants. After steady vibration is established in the 3D FSI simulations, 80 time frames of data in each case, which cover at least 2 vibration cycles (from 8 to 16 milli-seconds, depending on the frequency), are used for training. That is, at each chosen time frame, the instantaneous values of the six input variables and the three output variables represent a data point for machine learning.  
To calculate these input and output variables, we extract the flow velocity and pressure along the centerline of the 3D flow field. In total, there are 480 data points for all 6 cases together for training.  We will compare the output from the regression (i.e., the equations derived from the machine learning) with those provided for training.  To extend the validation, we consider Cases 7-10, in which the subglottal pressure is different and whose data have not been used for training, for further assessment. For all ten cases, we will also use the machine learning enhanced 1D flow model to replace the 3D flow and perform FSI simulations, and we will compare the vibration frequency, amplitude and phase delay of the vocal fold between 3D FSI and the simplified FSI model. The entire procedure is shown using a flow chart in a supplementary figure~\cite{figure_s1}.

\section{Results and discussion}

\subsection{Results from machine learning}
\label{sec:generate_data}

\begin{figure}
\centering
\epsfig{file=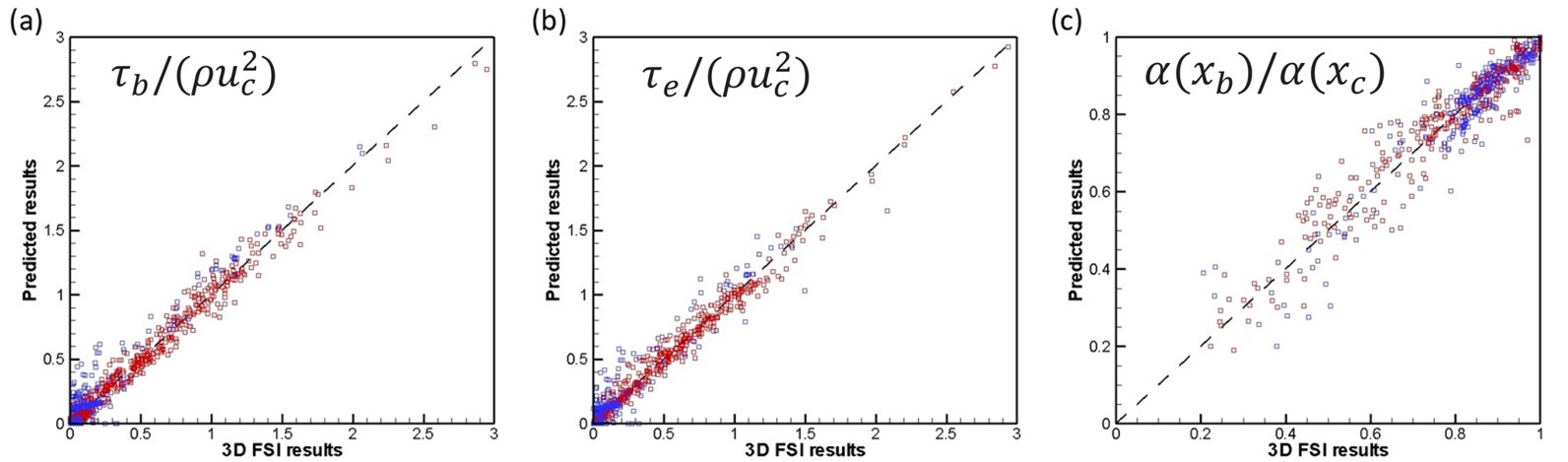, width=6.5in}  

\caption{ Comparison between 3D FSI results and the predicted results  from data regression.  Data in Cases 1-6 (red symbols) are used as training data; Data in Cases 7-10 (blue symbols) are only used for validation. (a) Pressure loss at the glottal exit, $\tau_b/(\rho u_c^2)$.  (b) pressure loss at $x_e$, $\tau_e/(\rho u_c^2)$. and (c) the entrance effect ratio at $x_b$, $\alpha(x_b)/\alpha(x_c)$. The dashed line represents the ideal fit.}
\label{f:ml-reg}
\end{figure}

After the training process, we obtain the explicit expressions of the output variables as defined in Section~\ref{s:ml-out}.  These expressions are given in the appendix A.  Results of data regression are shown in Fig.~\ref{f:ml-reg} for $\tau_b/(\rho u_c^2)$, $\tau_e/(\rho u_c^2)$, and $\alpha(x_b)/\alpha(x_c)$, where $u_c$ is the centerline velocity at the minimum section $x_c$.   In this figure, the $x$-axis represents the 3D FSI results of Cases 1 to 10 and the $y$-axis represents the predicted value based on the trained regression model, i.e., Eqns.~\eqref{e:taub} to \eqref{e:ml-ent}.  Red symbols represent training data from Cases 1-6, while blue symbols are the validation data from Cases 7-10. Ideally, the predicted value is equal to the 3D FSI value so that all data points would fall on the dashed line $y=x$ in the figure.  However, due to the error in the fitting process, the data are scattered around the line.  From this figure, we can see that both the training data and the validation data mostly cluster around $y=x$; thus, the predicted results from regression agree reasonably well with the 3D results. We calculate the mean error between the machine learning result and the benchmark result from the 3D simulations for each of these three output variables, i.e., $|\phi_{ML} - \phi_{3D}|$.  For $\tau_b/(\rho u_c^2)$, the mean error of the training data and mean error of the validation data are 0.029 and 0.033, respectively;  for $\tau_e/(\rho u_c^2)$, they are 0.057 and 0.086, respectively; and for $\alpha(x_b)/\alpha(x_c)$, they are 0.048 and 0.078, respectively. 

In Fig.~\ref{f:ml-reg}(a,b), we see that the loss coefficients in the glottis, $\tau_b/(\rho u_c^2)$ and $\tau_e/(\rho u_c^2)$, vary from nearly zero to nearly three.  The zero value means that there is no loss in the flow, while three represents a significant loss in the flow. Such great loss happens when the glottis is nearly closed and the flow speed becomes small, which is analogous to a closing mechanical valve in a pipe flow.

In Fig.~\ref{f:ml-reg}(c), we see that $\alpha(x_b)/\alpha(x_c)$ is clustered above 0.5, indicating that the entrance effect at the glottal exit is not necessarily significant at those time frames. Further examination shows that in those situations, the glottis has a small divergent angle or has a short divergent section, which does not create a strong entrance effect.  However, there are also a number of data points in this figure where $\alpha(x_b)/\alpha(x_c)$ is below 0.5 and is even close to 0.2. For those situations, the divergent section is typically long and/or has a large diverging angle, which causes early flow separation and strong entrance effect. The distribution of the data points in this figure thus covers a wide range of situations for the flow, which is preferred for the training purpose.   To further illustrate the data variation and validation of the machine learning, we have added a supplementary figure plotting the three output variables against time for the validation cases~\cite{figure_s2}.

Once the expressions for $\tau_b$, $\tau_c$, and $\alpha(x_b)/\alpha(x_c)$ are derived from machine learning, we can apply them in the 1D flow model.  In doing so, $\tau(x)$ and $\alpha(x)$ in the glottis take the assumed distribution as in Section~\ref{s:ml-out}.  That is, we assume a cubic function for $\tau(x)$ as shown in Fig.~\ref{f:ml-geo}(b), where $\tau=0$ and $\d \tau/\d x=0$ at $x_a$. When $x_c$ is very close to $x_b$, i.e., the glottis is almost purely convergent, the cubic interpolation may result in overshoot of the function. Thus, when $|x_c-x_b|$ is less than $H$, we disregard $\tau_e$ and switch the cubic function for $\tau(x)$ to the quadratic function with the same boundary conditions at $x_a$.  For the area correction coefficient $\alpha(x)$, we assume a quadratic function between $x_c$ and $x_b$, while requiring $\d\alpha/\d x =0$ at $x_b$, as shown in Fig.~\ref{f:ml-geo}(b). Between the glottal inlet $x_a$ and the minimum section $x_c$, the thickness of the boundary layer has only small change, so we assume that $\alpha(x)/\alpha(x_c) = 1$ in that region.

\begin{figure}
\centering
\epsfig{file=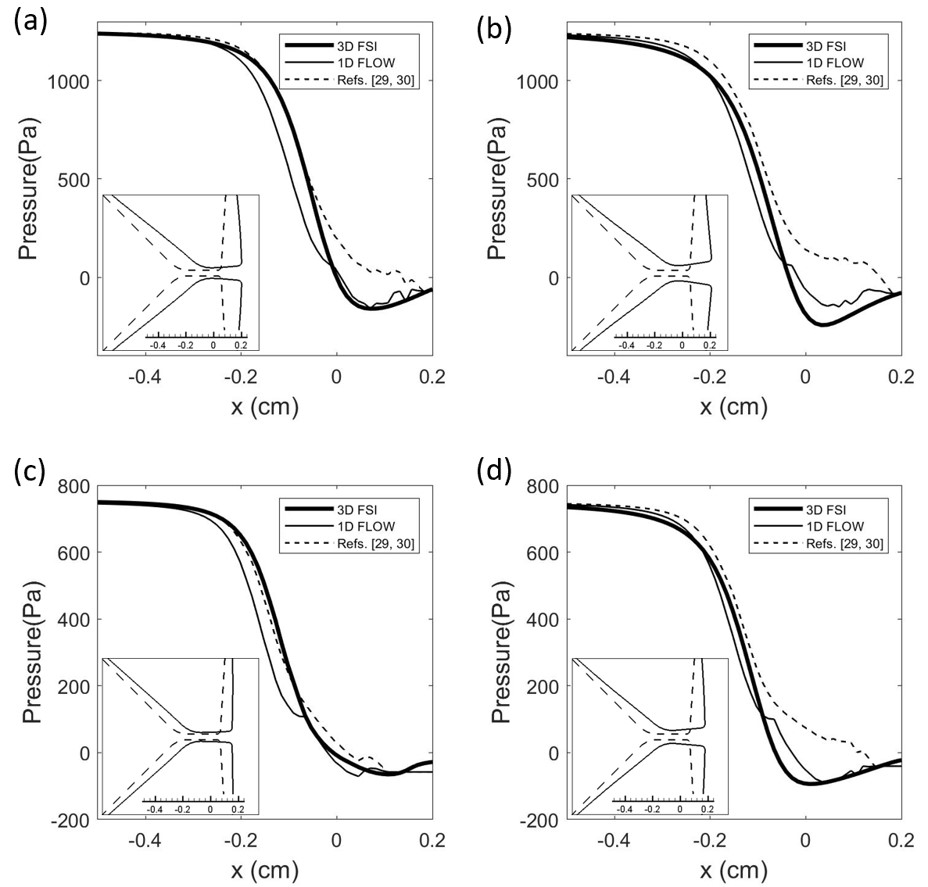, width=4.5in}  
\caption{ Comparison of pressure distribution along the centerline between 3D FSI and 1D flow model for small medial thickness $T$ cases. (a) Closing phase in Case 7, (b) opening phase in Case 7; (c) closing phase in Case 8, (d) opening phase in Case 8. The inset shows the corresponding vocal fold deformation (solid lines) from its initial configuration (dashed lines).}

\label{f:p-smallT}
\end{figure}

\begin{figure}
\centering

\epsfig{file=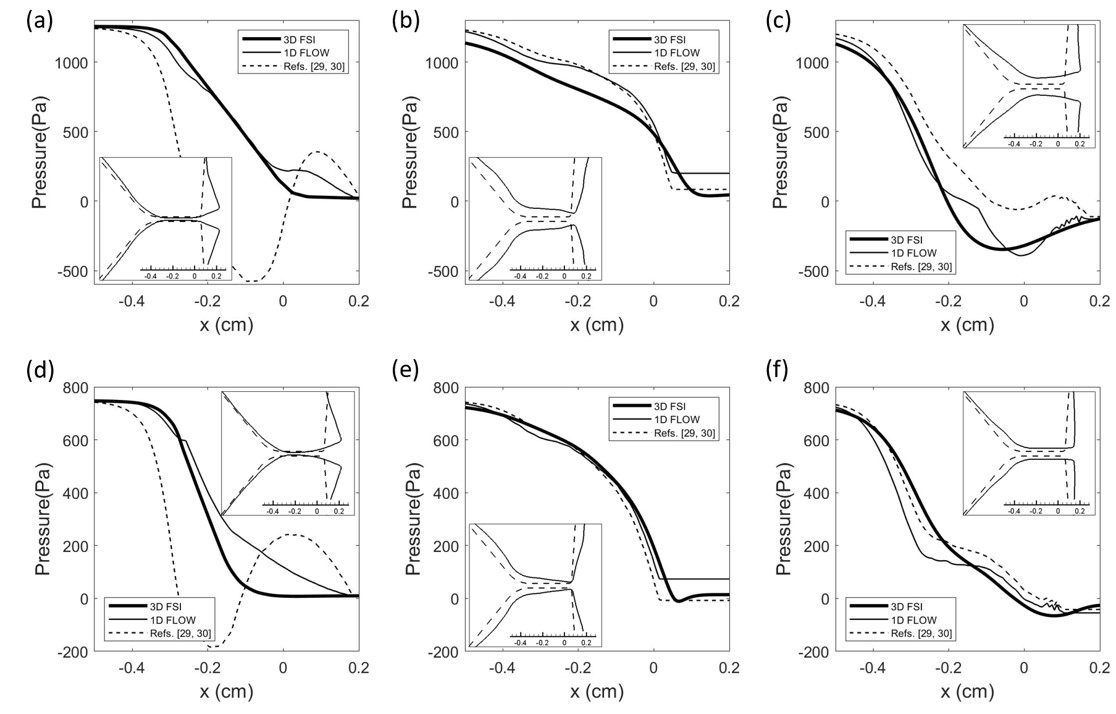, width=6.5in} 

\caption{ Comparison of pressure distribution  along the centerline between 3D FSI and 1D flow model for large medial thickness $T$ cases. (a) Closing phase, (b) opening phase, and (c)  maximum opening in Case 9; (d) closing phase, (e) opening phase, and (f)  maximum opening in Case 10. The inset is explained in Fig.~\ref{f:p-smallT}.}
\label{f:p-largeT}
\end{figure}

 To verify the 1D flow model enhanced by machine learning, we first compared the pressure distribution predicted by this model against 3D results generated from cases that are not used in the training process.  More specifically, we use the instantaneous glottal shape obtained from the 3D FSI simulations of Cases 7 to 10 and calculate the pressure distribution using the 1D flow model in Eq.~\eqref{e:mom}. Then, the result is compared with the pressure at the centerline extracted from the 3D flow field.  

Figure~\ref{f:p-smallT} shows such comparison of pressure distribution at vocal fold opening and closing phases from Cases 7 and 8, which have the same medial thickness of $T=1.75$ mm but different $P_{sub}$.  It can be seen that the pressure produced by the 1D flow model is close to that from the 3D simulation for different $P_{sub}$ values, including the negative pressure during the opening phase when the glottis is of divergent shape. From our previous study~\cite{li2019reduced}, correct prediction of the negative pressure in this case of small medial thickness is important; otherwise, the vocal fold may exhibit a different vibration mode that is associated with the first eigenmode of the tissue structure~\cite{li2019reduced}.  

Figure~\ref{f:p-largeT} shows comparison of the pressure distribution for Cases 9 and 10, which have the same medial thickness of $T=3.50$ mm but  different $P_{sub}$.  In these cases, the glottis is relatively long and may form diverging, converging, and converging-diverging shapes at different vibration phases.  From our previous study~\cite{li2019reduced}, if a Bernoulli equation based model is used, then an inappropriate setting of the location for flow separation may lead to exceedingly negative pressure that might destabilize the FSI simulation.  In the present study, the 1D flow model correctly predicts the negative pressure zone (e.g., Fig.~\ref{f:p-largeT}(c) and (f)) and captures the pressure reasonably well at different glottal shapes.

In Figs. \ref{f:p-smallT} and \ref{f:p-largeT}, we also include the pressure distributions calculated by the untrained 1D flow model with {\it ad hoc} assumptions in pressure loss and entrance effect~\cite{li2019reduced, li2019reduced2}. Overall, the current 1D flow model trained by machine learning achieves better accuracy as compared with those reference results.

\begin{figure}
\centering

\includegraphics[width=6.5in]{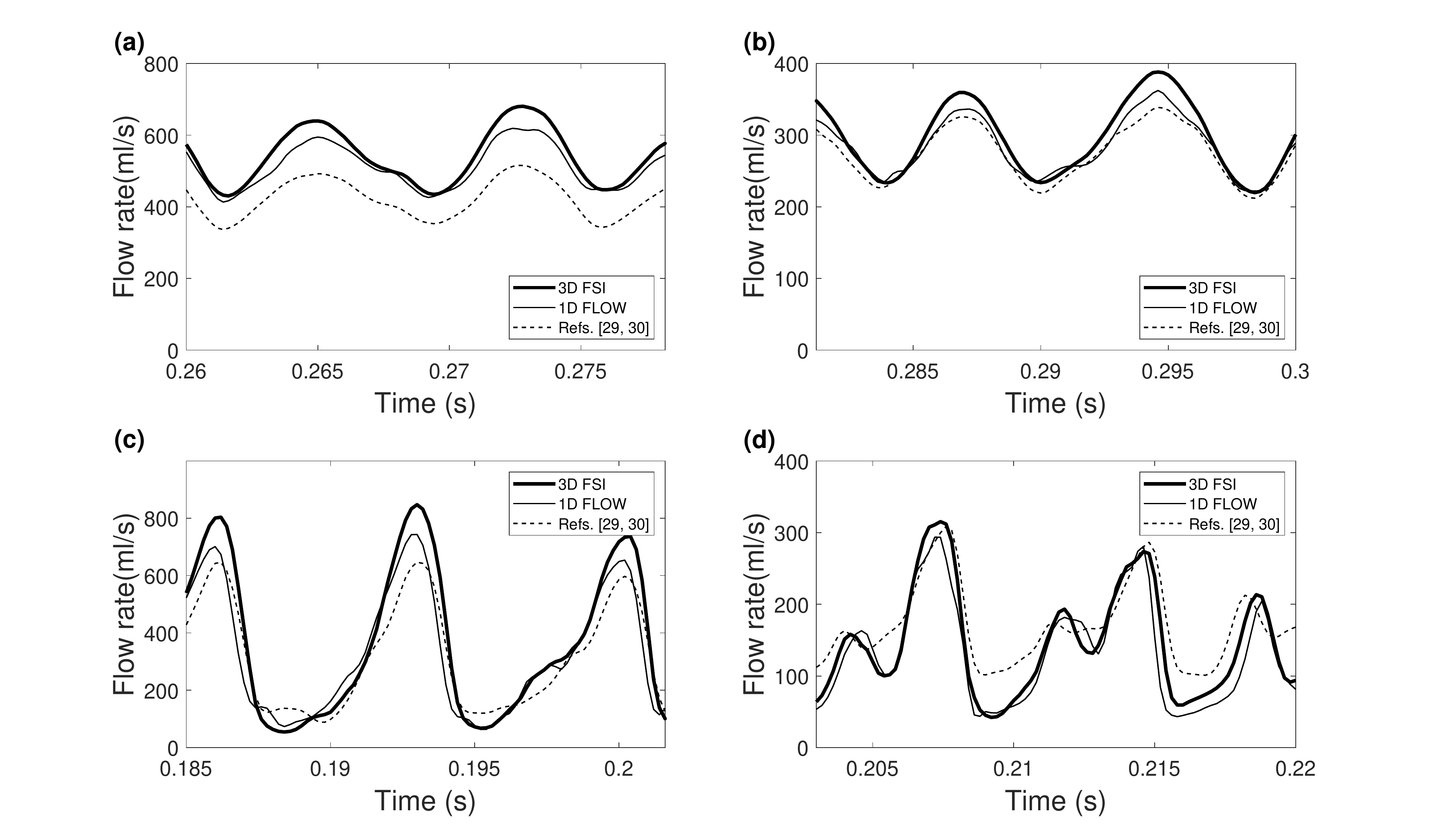}

\caption{ Comparison of the volume flow rate between 3D FSI and 1D flow model for (a) Case 7, (b) Case 8, (c)  Case 9, (d) Case 10, which are not in the training data set.}
\label{f:flowrate}
\end{figure}

 Besides the pressure distribution, the flow rate is also important for the glottal airflow. Figure~\ref{f:flowrate} compares the volume flow rate between 3D FSI and 1D flow model for Cases 7 to 10 that are not used in the training process. Similar to the pressure distribution, we calculate the volume flow rate in 1D flow model using the same glottal shape from the 3D simulation.  Two representative cycles are selected for each case comparison.  In Cases 9 and 10 where the medial thickness is larger, the vocal fold has better closure, and thus the flow rate has greater oscillations and is reduced to nearly zero at closure.  On the other hand, in Cases 7 and 8 where the medial thickness is smaller, the vocal fold maintains a significant gap at the closing phase, which leads to a high flow rate and low-magnitude oscillations during vibration. In all cases under consideration, the flow rate from 1D flow model agrees well with 3D FSI result.

As a reference, we also include the volume flow rate in Cases 7-10   calculated using the untrained 1D flow model~\cite{li2019reduced, li2019reduced2}. Comparing with the reference result, the flow rate from the present model shows significantly better agreement with the 3D FSI result. 

Therefore, the present 1D flow model provides reliable predication for the volume flow rate during vocal fold vibration.

\subsection{Application in the FSI of idealized vocal fold models}

\begin{figure}
\centering

\epsfig{file=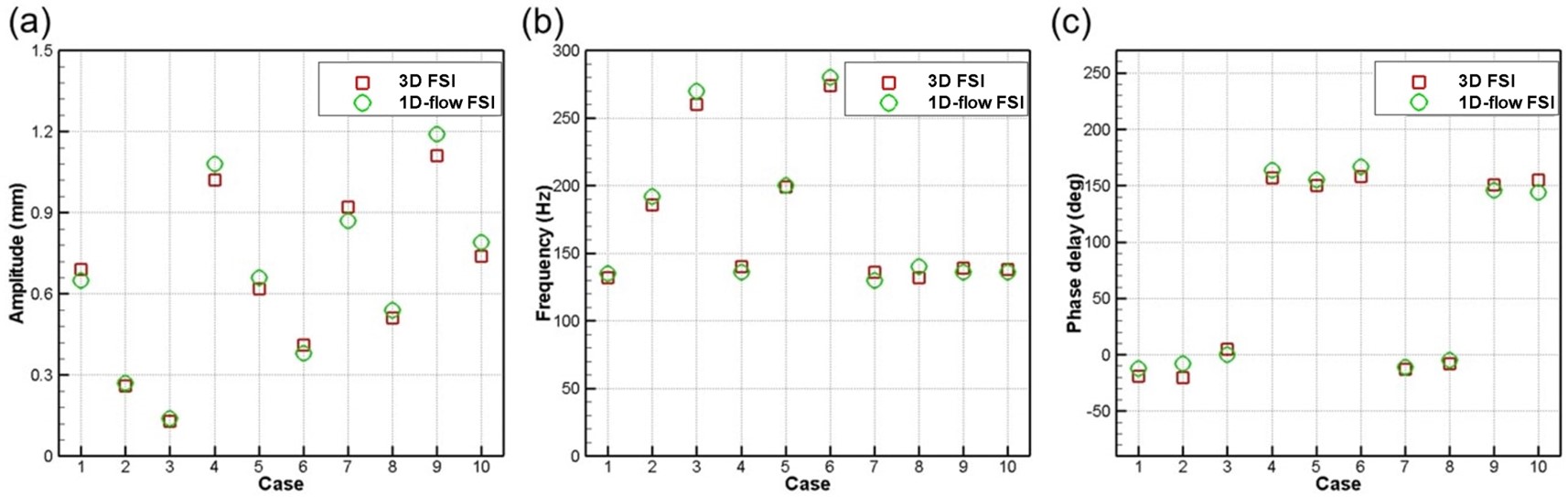, width=6.5in}  

\caption{ Comparison between 3D FSI and reduced-order model: (a) amplitude, (b) frequency, (c) phase delay. }
\label{f:ideal}
\end{figure}

After verifying the 1D flow model for flow calculation only, we then apply it in the FSI simulation by coupling it with the 3D idealized vocal fold model described in Section~\ref{sec:generate_data}.  We compare the vibration characteristics from this 1D-flow/3D-solid hybrid FSI simulation with those from the full 3D FSI simulation.  For this comparison, all ten cases in Table~\ref{t:ml-cases} are considered, which include variations in the medial thickness, stiffness properties, and subglottal pressure. As mentioned before, Cases 7 to 10 were not used in the training data but included here as additional validation.

We use the vibration amplitude, frequency, and phase delay for quantitative comparison between the two sets of simulations.  The vibration amplitude is defined as the maximum $y$-displacement of the vocal fold measured at the glottal exit in a cycle. From Case 1 to Case 3, the tissue stiffness increases while the other parameters are the same.  As shown in Fig.~\ref{f:ideal}(a), the vibration amplitude decreases with increasing tissue stiffness from Case 1 to 3, which have a smaller medial thickness.  Similar trend can be seen from Case 4 to 6, which have a larger medial thickness.  Cases 7, 1 and 8 have different subglottal pressures from  high to low. Correspondingly, their vibration amplitude shows a decrease in the same order. A similar result can be seen for Cases 9, 4, and 10 in which $P_{sub}$ is  decreased. In all ten cases, the two sets of FSI simulations produce closely agreeable results.

 The comparison of the vibration frequency is shown in Fig.~\ref{f:ideal} (b). The second-eigenmode type vibration~\cite{luo2009analysis} is established in all the cases, where the vocal fold oscillation is primarily in the lateral or $y$-direction, and this mode is captured by the 1D-flow FSI simulation. Thus, the frequency predicted by this hybrid FSI matches that by the full 3D FSI.  From the figure, the effect of the tissue stiffness on the vibration frequency is clear, e.g., from Case 1 to 3, and from Case 4 to 6, where an increase of the tissue stiffness leads to increase of the vibration frequency.

The phase delay is calculated using the temporal difference between the glottal inlet and the exit in the mid $xy$-plane in terms of the displacement. Good agreement is again achieved in the comparison between the two sets of simulations for all cases.  For this quantity, the most influential parameter is the medial thickness, as the longer glottis would create greater phase difference between the glottal inlet and the exit and thus lead to a more pronounced mucosal wave along the glottis.


To compare the performance of the present flow model with that of the untrained flow model, we include the results from a previous study~\cite{li2019reduced,li2019reduced2}.  This comparison is shown in {\blu Table}~\ref{t:FSI-comp}.  Only one case each from the large $T$ and small $T$ is shown here for brevity. We also include the average differences for the frequency, amplitude and phase delay from Cases 1-10. {\blu If we only consider Cases 7-10, the average error is 3.3\%, 6.3\% and 5.2$^\circ$, for the frequency, amplitude, and phase delay, respectively, which is consistent with the overall average in Table~\ref{t:FSI-comp}.} From the comparison, it can be seen that the trained flow model leads to better accuracy in the predication of the vibration than the previous untrained model. 

\begin{table}[]
\begin{center}
    \caption{ 1D-flow FSI results compared with 3D FSI in terms of vibration frequency $f$, amplitude $d$, and phase delay $\phi$. Results using the untrained model in Ref.~\cite{li2019reduced} are also included.}
\begin{tabular}{ |c|c|c|c|c|c|c|c| } 
\hline 
 & Model & f (Hz) & difference & d (mm) & difference & $\phi$ ($^\circ)$  & difference ($^\circ$)\\
\cline{1-8}
\multirow{3}{4em}{Case 1} & 3D FSI & 132 & -& 0.69 & - & -19 & -\\ 
\cline{2-8}
& 1D-flow FSI & 135 & 2.3\% & 0.65 & 5.8\% & -15 & 4\\ 
\cline{2-8}
& Ref \cite{li2019reduced} & 144 & 9.1\% & 0.58 & 15.9\% & -15 & 4\\ 
\hline
\multirow{3}{4em}{Case 4} & 3D FSI & 140 & - & 1.02 & - & 157 & -\\ 
\cline{2-8}
& 1D-flow FSI & 136 & 2.9\% & 1.04 & 2.0\% & 159 & 2\\ 
\cline{2-8}
& Ref \cite{li2019reduced} & 144 & 2.9\% & 1.00 & 2.0\% & 161 & 4\\ 
\hline

\multirow{2}{6em}{Avg. error (Cases 1-10)} & 1D-flow FSI & - & 2.3\% & - & 5.3\% & - & 6\\ 
\cline{2-8}
& Ref \cite{li2019reduced,li2019reduced2} & - & 3.9\% & - & 9.1\% & - & 11\\ 
\hline
\end{tabular}
\label{t:FSI-comp}
\end{center}
\end{table}

\subsection{Application in the subject-specific vocal fold models}
\label{sec-subject}

\begin{figure}
\centering
\epsfig{file=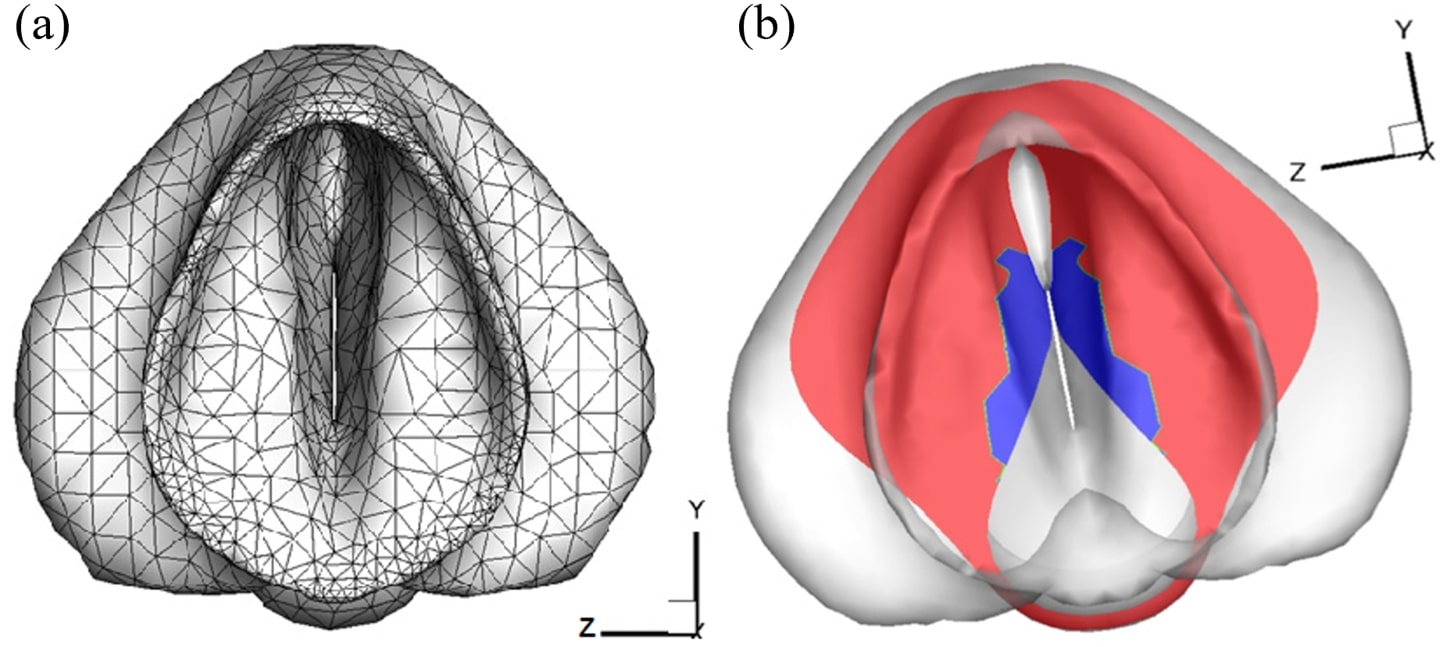, width=5.0in}

\caption{Subject-specific vocal fold model: (a) reconstructed larynx geometry from a superior view; (b) profiles of the vocal fold cover (blue) and body (red) segmented from the MRI scan.}
\label{f:pmodel}
\end{figure}

Other than the idealized vocal fold geometry, we also apply the 1D flow model in the FSI simulation of the subject-specific vocal fold models that were generated based on 3D scan of rabbits' larynx (Fig.~\ref{f:pmodel}).  In the present study, we utilize the vocal fold models created previously in Chang et al.~\cite{chang2015subject} and will validate the simulation results against the experimental data of evoked {\it in-vivo} phonation. The same models were also used in Chen~\etal~\cite{li2019reduced2} to validate the 1D flow model without machine learning.  Readers are referred to Chang et al.~\cite{chang2015subject} for the details how these anatomical models were created and how the {\it in-vivo} measurement of the vocal fold vibration was conducted.  Only a brief summary is given here to provide the context.

In the experiment~\cite{novaleski2016nonstimulated}, live rabbits were used in the study; their vocal fold was surgically sutured to achieve adduction, and phonation was evoked by introducing pressurized air from their trachea.  High-speed videos of the vocal fold vibration were taken during the experiment, which provide the vibration frequency, magnitude, and waveform as the validation data for our current study.  

After the phonation experiment, the rabbit larynx was excised and high-resolution MRI was performed to obtain details of morphology of the vocal fold while the vocal fold maintained the adducted phonatory position.  The 3D anatomical vocal fold model was generated for each of the five samples after manual segmentation from the MRI data and surface mesh reduction/smoothing~\cite{chang2015subject}.  Furthermore, the tissue properties were estimated in that study through simulations, and these properties of the five samples can be found in our previous publications~\cite{chang2015subject,li2019reduced2}.

\begin{figure}
\centering

\epsfig{file=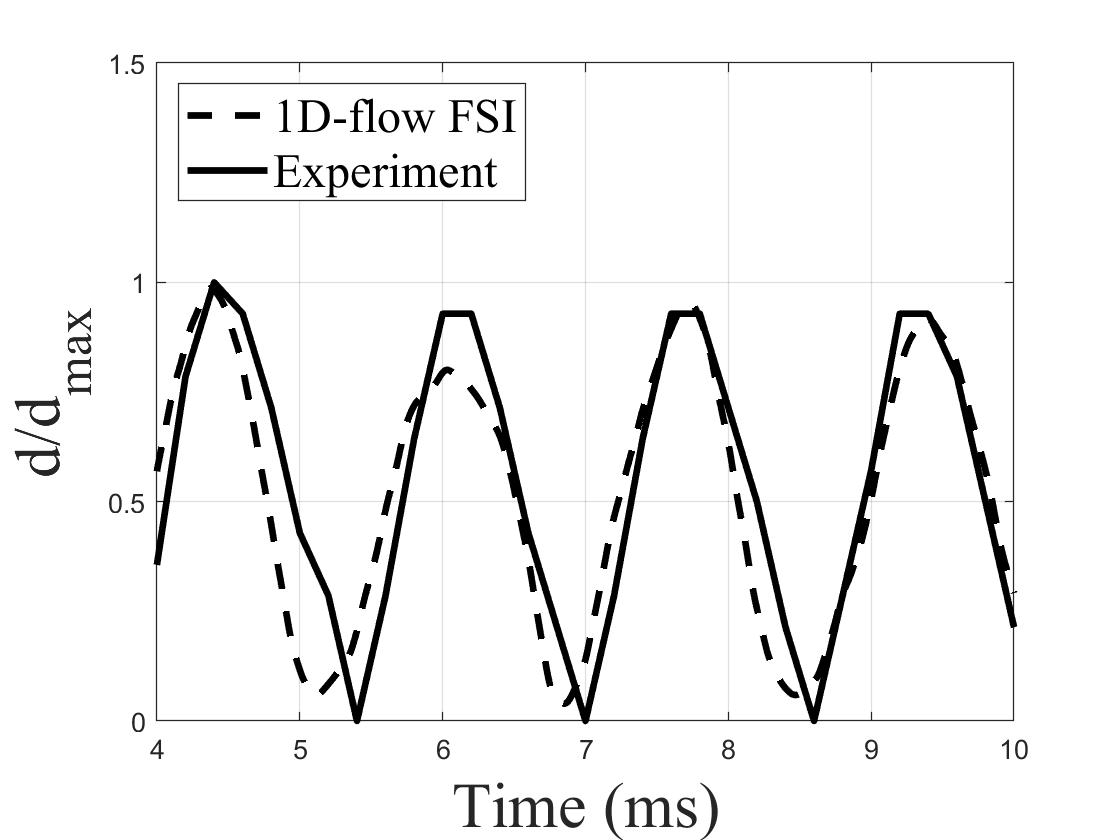, width=4.0in}  
\caption{Waveforms of the normalized glottal gap width from the {\it in-vivo}  phonation experiment and the FSI simulation for the rabbit sample R1. }
\label{f:waveform}
\end{figure}

To perform the hybrid 1D-flow/3D-solid FSI simulation, we couple the FEM representation of each anatomical vocal fold with the present trained 1D flow model.  The subglottal pressure in each case is obtained from the experiment and is in the range of 0.72 to 1.05 kPa~\cite{chang2015subject}. Figure~\ref{f:waveform} shows a comparison of the normalized glottal gap width in a sequence of vocal fold oscillations between the FSI simulation and the experiment. Because the high-speed imaging does not provide a length scale, we use the normalized gap width, $d/d_{\max}$, for comparison, where $d$ is the gap width of the glottis measured at mid-section and $d_{max}$ is its peak value.  From this figure, it can be seen that the waveform obtained from the simulation agrees generally well with the experiment. {\blu In modeling the vocal fold contact, we maintain a minimum glottal gap of 0.02 mm between the sides for the flow~\cite{luo2008immersed}. Thus, the waveform from the FSI simulation does not have full closure.}  For quantitative comparison, we further compute the normalized root-mean-square (r.m.s.) error of the waveform between the simulation data and experiment for each sample.  The result is listed for all five samples in {\blu Table}~\ref{t:rms}, which shows that the error is within 15\% for all cases.  As shown in the table, these r.m.s errors are lower than the results in the previous work, where the untrained flow model was used~\cite{li2019reduced2}. Therefore, the trained flow model provides improved results and reasonable prediction of the vibration for these subject-specific models. 

\begin{table}
\begin{center}
\caption{The normalized r.m.s. error of the gap width waveform for each rabbit sample.  Results from a previous study~\cite{li2019reduced2} with the untrained flow model is also included.}

\begin{tabular}{ |c|c|c|c|c|c| } 
\hline
Sample & R1 & R2 & R3 & R4 & R5\\
\hline
Error & 13.7$\%$ & 10.3$\%$ & 12.7$\%$ & 14.5$\%$ & 14.0$\%$\\
\hline
Error~\cite{li2019reduced2}& 14.3$\%$ & 11.3$\%$ & 15.3$\%$ & 15.9$\%$ & 16.9$\%$ \\ 
\hline
\end{tabular}
\label{t:rms}
\end{center}
\end{table}

\begin{figure}
\centering
\epsfig{file=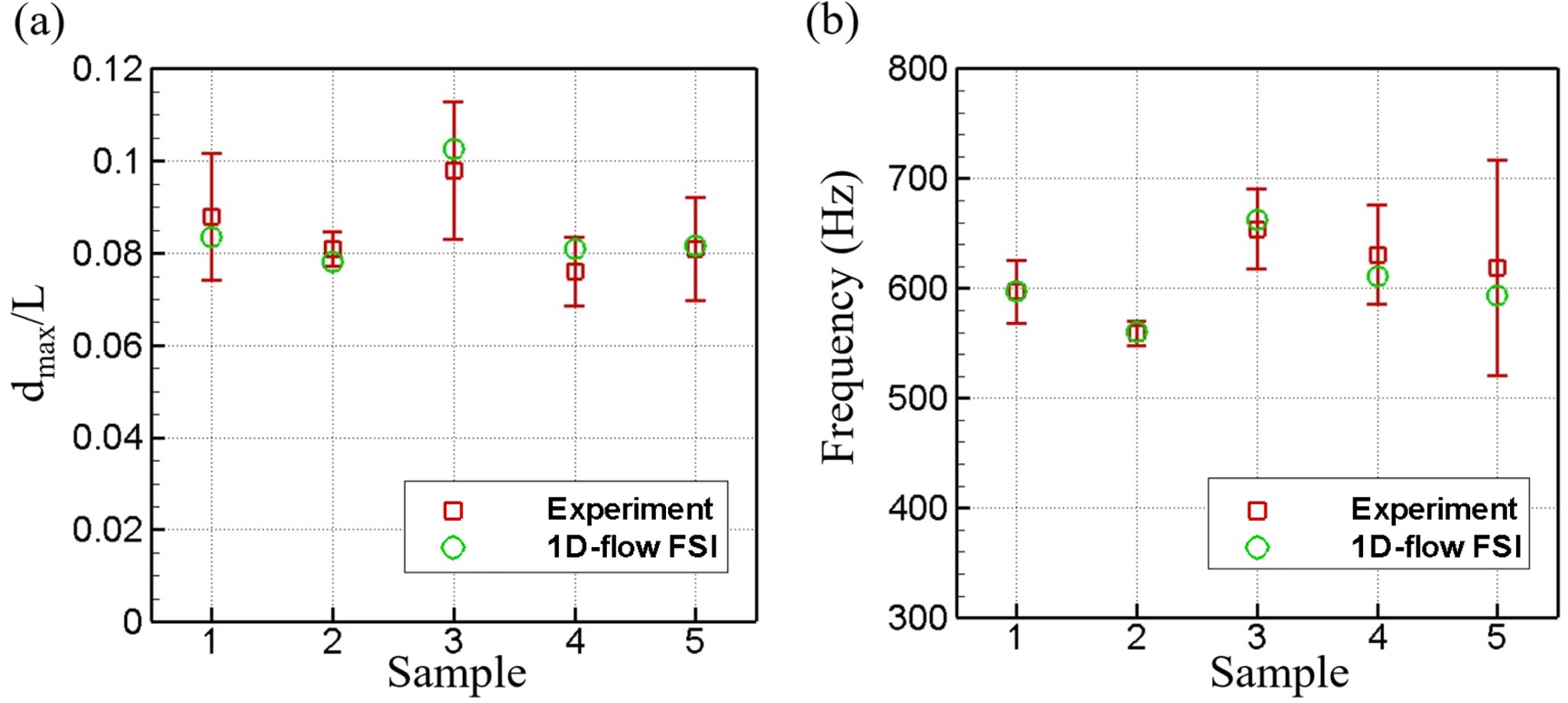, width=6.0in} 

\caption{Frequency and amplitude comparison between the experimental and numerical results for five rabbit samples. }
\label{f:patient}
\end{figure}

Figure~\ref{f:patient} further shows a quantitative comparison between the experiment and the FSI simulation for all five samples in terms of the vibration frequency and normalized amplitude, $d_{max}/L$, where $L$ is the vocal fold length.  {\blu In the experiment, each sample had three trials to analyze the standard
deviations ~\cite{novaleski2016nonstimulated,chang2015subject}.} From this figure, both the frequency and the amplitude from the simulation fall within the range of the experimental data for all the five samples despite significant variations among the individual subjects.  This result again confirms the performance of the present 1D flow model in the FSI simulation.

\subsection{Discussion}
In contrast with 3D computational fluid dynamics models that employ extensive computing resources and require substantially longer simulation time, the drastically simplified flow models such as the present 1D model offers much faster turnaround and may be used in conjunction with the 3D models as a complementary fashion for model-based prediction.  The performance of such models could be evaluated in terms of their accuracy, robustness, and required information during practical implementation.  

Using a similar set of partial differential equations based on momentum and mass conservation, the present 1D flow model retains the viscous and entrance effects of the flow model from the previous studies of Luo and coworkers~\cite{li2019reduced,li2019reduced2} and hence offers similar advantages shown therein in comparison with the traditional Bernoulli based models.  In the present model, we have incorporated machine learning to generalize the pressure loss and entrance effect in the glottis. Those effects were only assumed by ad hoc manners previously in the untrained model~\cite{li2019reduced,li2019reduced2} and may need adjustment in practical use.  For example, previously the shear stress related to flow separation, $\tau_{\chi}$, is modelled as 
$\tau_{\chi} = \frac{A}{s} (\,1-\chi)\, \rho u \frac{\partial u}{\partial x}$, where $s$ is the perimeter around the cross section, $A$ is the effective cross section area, and $0\le \chi \le 1$ is a constant representing pressure recovery (see Eq.~(2) in Ref.~\cite{li2019reduced2}). In its application, the value of $\chi$ has to be adjusted empirically for a divergent channel in order to achieve matching results to the 3D model. In addition, the area correctional coefficient, $\alpha(x)$, has an adjustable constant $C_1$ in Eq.~(4) of Ref.~\cite{li2019reduced2}.  In the present study, the parameters in the shear stress and the area correctional coefficient have been expressed in explicit functions of the Reynolds number and a few geometrical parameters describing the instantaneous shape of the glottis through a data regression procedure and thus has better capability of generalization.    

Using the idealized vocal fold geometry, we have demonstrated that the new flow model provides better prediction of the pressure distribution and flow rate than the previous untrained model (Section 3.1).  When applied to the FSI simulation, the new flow model leads to clearly more accurate predication of the vibration frequency, amplitude, and phase delay in the vocal fold dynamics (Section 3.2).  However, when it is applied to the subject-specific vocal fold geometries, the new flow model offers only small or limited improvement as compared with the previous untrained model (Section 3.3), {\blu where the normalized r.m.s for all samples ranges from 10\% to 15\%.} The reason for this limitation is mostly likely due to presence of many other uncertainties related to this type of subject-specific models, e.g., quality of the MRI data, the segmentation errors, assumption of the tissue properties, as well as the experiment itself, which start to become predominant factors over the numerical model's own error. In that case, improving the numerical model alone will not further increase the overall accuracy. 

Despite being unable to substantially improve the accuracy in the subject-specific cases, the present flow model is still advantageous as compared with the similar existing models. We emphasize that from the idealized geometries (including significant variations in the medial thickness, subglottal pressure, and the tissue stiffness) to the anatomical geometries, the current 1D flow model uses exactly the same pressure loss and entrance effect functions that are derived from machine learning, and there is no need to make any parameter adjustment. This feature of robustness, the overall improved accuracy in all the tests, as well as the fact the present model does not require any additional input information, indicate that the present flow model has significant better performance than the previous untrained model~\cite{li2019reduced,li2019reduced2}.

\section{Conclusion}

In this study, we have presented a new 1D flow model for glottal airflow that is based on the viscous flow assumption.   As compared with similar models in the previous studies, in the current flow model we derive the pressure loss and the entrance effect using a machine learning approach and express them as explicit functions of the Reynolds number and the parameters describing the characteristic shape of the glottis at any instantaneous moment.  Unlike previous models in which the parameters need to be modified ad hoc for different cases such as the vocal fold geometry, the present machine-trained model can be used for more general situations without the need to modify its parameters. We have tested the performance of this 1D flow model in three scenarios.  First, we use this flow model to calculate the pressure distribution and the volume flow rate using the glottal configuration from the 3D FSI simulations not included in the training process. The results agree well with those directly from the 3D simulations, and they are significantly better than those from the previous untrained model. Second, the 1D flow model is coupled with the 3D idealized vocal fold geometry to perform the hybrid 1D-flow/3D-solid FSI simulation, and the results show that the vibration characteristics match the full 3D FSI simulation results significantly better than the untrained model in terms of the vibration frequency, amplitude, and phase delay.  Third, we applied the 1D flow model to the subject-specific vocal fold models constructed from the rabbit larynx, and the FSI simulation results are compared against the previous {\it in-vivo} experimental data. Even though in this case the improvement is limited likely due to the presence of uncertainties, the new model  achieves the accuracy performance without the need to adjust its parameters.

In summary, we conclude that the present 1D glottal airflow model enhanced by machine learning is more accurate and robust than the similar models in the previous studies and could be useful for efficient modeling of vocal fold dynamics, e.g., estimate of the unknown tissue properties of an individual subject's vocal fold using model-based simulations, or design optimization of the surgical implant inserted into a paralyzed vocal fold.


\appendix
\section{Expressions from sparse regression}

The following formulae are the expressions derived from the SINDy method for the pressure losses $\tau_b$, $\tau_e$, and the area correction coefficient $\alpha(x)$ at $x_b$.
\begin{eqnarray}
\frac{\tau_b}{\rho u_c^2} &=& 5.1282\xa 
 -0.1902\xb 
 -2.2354\xf 
 +3.3258\xe 
 +0.2611\xc 
 -0.0553\xd \nn\\
 &-& 0.1955 \xa^2
 -0.9508 \xa\xe
 -0.2648 \xa\xc  
 +0.0606 \xb \xc 
 +0.0759\xg \nn\\
 &+& 0.2523  \xf  \xc 
 -0.2052 \xe^2 
 -0.2450 \xe\xc 
 +0.2834 \xc^2 \nn\\
 &+& 0.0680 \xa \xe^2 
 -0.1170 \xa \xc^2 
 -0.0361 \xf \xc^2 
 \label{e:taub}
\end{eqnarray}
%
\begin{eqnarray}
\frac{\tau_e}{\rho u_c^2} &=& 20.8293 \xa 
 -3.4297\xb 
 +0.0784\xf 
 -0.2519\xe
 -1.8603\xc 
 -0.0637\xd  \nn\\
 &-& 2.0506\xa^2
 +0.2090  \xa   \xb
 -1.9726  \xa   \xf
 -1.7743  \xa   \xe
 -1.1751   \xa   \xc  \nn\\
 &+& 0.0292   \xb   \xf
  +0.8267   \xb   \xe
  +0.5027   \xb   \xc
  +0.2645   \xe   \xc 
  + 0.5762   \xc^2 \nn\\
  &+& 0.0636   \xa^3
  +0.1096   \xa^2   \xf
  +0.0548   \xa^2  \xe
  -0.0316   \xa   \xb   \xe
  +0.0593   \xa   \xg
  +0.1785   \xa   \xf   \xc \nn\\
  &+& 0.1384   \xa   \xe^2
  -0.2707   \xa   \xc^2
  -0.0544   \xb   \xe^2
  -0.0796   \xb   \xe   \xc
  -0.0300   \xf   \xc^2 
\label{e:taue}
\end{eqnarray}
%
\begin{eqnarray}
\frac{\alpha(x_b)}{\alpha(x_c)}  &=&  1.6652\xa 
 -3.4873  \xb 
 +0.0285  \xf 
 +1.1238  \xe 
 +0.3850  \xc  \nn\\
 &-& 0.8358   \xa^2
 +0.9378   \xa   \xb 
 -0.0559   \xa   \xf 
 -1.1604   \xa   \xe 
 -0.5005   \xa   \xc \nn\\ 
 &+& 0.6961   \xb   \xe
 +0.8935   \xb   \xc
 -0.0344   \xf   \xd
 -0.0938   \xe^2
 -0.1952   \xe   \xc
 +0.0449   \xe   \xd  \nn\\
 &+& 0.0716   \xa^3
 -0.0551   \xa^2   \xb
 +0.1134   \xa^2   \xe
 +0.0926   \xa^2   \xc
 -0.0788   \xa   \xb   \xe
 -0.1336   \xa   \xb   \xc \nn\\
 &+& 0.0373   \xa   \xf   \xd 
 + 0.0900   \xa   \xe^2
 +0.2041   \xa   \xe   \xc
 -0.0504   \xa   \xe   \xd
 -0.0353   \xb   \xe^2 \nn\\
 &-& 0.0896   \xb   \xe   \xc
 -0.0565   \xb   \xc^2
\label{e:ml-ent}
\end{eqnarray}

\vskip0.2in
\noindent
{\bf Acknowledgement:}
This research was supported by an NIH grant 5 R01 DC016236 03 from the National Institute of Deafness and Other Communication Disorders (NIDCD).


\vskip0.2in
\bibliographystyle{elsarticle-num}

\end{document}